%% file: main.tex
\documentclass[reprint,aps,pre]{revtex4-2}
\usepackage{silence}
\usepackage{graphicx}
\usepackage{amsmath}
\usepackage{amsthm}
\usepackage{amssymb}
\usepackage{lineno}
\usepackage{rotating}
\usepackage{epsfig}
\usepackage{dcolumn}
\usepackage{bm}
\usepackage{mathptmx}
\usepackage{caption}
\usepackage{setspace}
\usepackage{threeparttable}
\usepackage{booktabs}
\usepackage{tabularx}
\usepackage{float}
\usepackage{chngcntr}
\usepackage{comment}
\usepackage{verbatim}
\usepackage{sidecap}
\usepackage{epstopdf}
\usepackage{soul}
\usepackage{xcolor}
\usepackage{url}
\usepackage{braket}
\usepackage{array}
\usepackage{subcaption}
\usepackage{anyfontsize}
\usepackage{natbib}

\graphicspath{./}
\usepackage[margin=1in]{geometry}

\begin{document}
\def  \degree  {{$^\circ$}}
\def  \water  {{\mathrm{H}_2\mathrm{O}}}

\title{
Structural and Energetic Stability of the Lowest Equilibrium Structures of Water Clusters}

\author{Vishwa K. Bhatt}
\email{vishwa\_bhatt@yahoo.com}
\affiliation{
        Department of Physics (Autonomous),
        University of Mumbai, Kalina Campus, Santacruz (East),
        Mumbai-400 098, India.
	}
 
\author{Sajeev S. Chacko}
\affiliation{
        Department of Physics (Autonomous),
        University of Mumbai, Kalina Campus, Santacruz (East),
        Mumbai-400 098, India.
	}

\author{Nitinkumar M. Bijewar}
\affiliation{
        Department of Physics (Autonomous),
        University of Mumbai, Kalina Campus, Santacruz (East),
        Mumbai-400 098, India.
	}
 
\author{Balasaheb J. Nagare}
\thanks{corresponding author}
\email{bjnagare@gmail.com}
\affiliation{
        Department of Physics (Autonomous),
        University of Mumbai, Kalina Campus, Santacruz (East),
        Mumbai-400 098, India.
	}
\date{\today}

\begin{abstract}
In the present work, the low-lying structures of 20 different sized water clusters are extensively searched using the artificial bee colony algorithm with TIP4P classical force field. To obtain the lowest equilibrium geometries, we select the 10 lowest configurations for further minimization using density functional theory. The resulting structures are lower in energy than previously reported results. The structural and energetic stability of these clusters are studied using various descriptors such as binding energy, ionization potentials, fragmentation energy, first and second energy difference, vibrational and optical spectra. The energetic analysis shows that clusters with N~$=4, 8, 12, 14, 16$ and 19 are more stable. The analysis of fragmentation energies also supports these findings. Our calculations show that non covalent interactions play a significant role in stabilizing the water clusters. The infrared spectra of water clusters display three distinct bands: intermolecular O...H vibrations (23–1191 cm$^{-1}$), intramolecular H-O-H bending (1600–1741 cm$^{-1}$) and O-H stretching (3229–3877 cm$^{-1}$). The strongest intensity is observed in the low-frequency symmetric stretching modes, along with a noticeable red-shift in the stretching vibrations. The optical band gap ranges from 7.14 eV to 8.17 eV and lies in the ultraviolet region. The absorption spectra also show line broadening for clusters with n $\geq$ 10, resulting in an increase in spectral lines. Interestingly, only the stable clusters exhibit maximum oscillator strength, with the first excitation in all cases corresponding to a $\pi \rightarrow \sigma^*$ transition.
\end{abstract}

\maketitle

\section{Introduction}
\label{introduction}
Every aspect of our daily lives is influenced by water due to its various unique properties and characteristics, suitable for the origin of life and its abundance in nature. It has been of prime importance due to its broad range of physicochemical properties paving the way towards various technological processes~\cite{article} such as corrosion, lubrication, heterogeneous catalysis and electrochemistry. It further helps to understand how different biological and geological processes are carried out, its major role in the operation of hydrogen fuel cells, as well as in advancing technological development at the nanoscale~\cite{fayer}.
	
Water clusters are found everywhere in nature, in the form of clouds, atmosphere, oceans etc~\cite{doi:10.1063/1.5031083}. These are formed due to various reasons like the presence of dust particles in the atmosphere, various micro- and nano-sized impurities, bio-waste, and solutes in the seawater. Depending upon these factors and various other factors like temperature, pressure etc around them, and the dimensions and the chemical properties of the surface or volume surrounding them, different-sized water clusters are formed, e.g. dimers, trimers etc. to clusters with $n$-molecules. These clusters with $n$-number of molecules have different bond lengths and bond angles, which lead to different physical and chemical processes observed in nature. It is found that the absorption of UV radiation in the atmosphere has led to many phenomena like water splitting, making available hydrogen and oxygen, OH$^-$, H$^+$ and other species~\cite{doi:10.1063/1.5031083}. Recently, it has been observed that the smallest size of water cluster~\cite{kenneth19} in which ice can form, is of 90 water molecules, where they have also studied the characteristic hydrogen bonding of ice-I between 90 water olecules using theoretical methods and experimental infrared spectroscopy, and found that in the range of clusters between 90 to 150 molecules, at a little below freezing temperature, there is a coexistence of crystalline and amorphous clusters and they exhibit hetero-phasic oscillations in time~\cite{moberg2019end}. It has also been seen that in biological systems, water molecules form an unlimited hydrogen-bonded network with structured clustering~\cite{chaplin01} and influences chain folding~\cite{levy04}, internal dynamics, conformational stability, binding specificity and catalysis~\cite{pocker2000}.
    
It has been observed that despite its simple molecular structure, water clusters provide a challenging problem to study since they involve a rather complicated interplay of various interactions, including vibration, bending and electrostatic interactions as well as Lennard-Jones (LJ) type interactions and complicated potential energy surfaces~\cite{hammer04}.  The study of the finite-size water clusters provides more accurate models of water, which may lead to an improved understanding of  bulk water. Water clusters can be considered as a bridge between the gas and the condensed phases, and  therefore, the evolution towards condensed phase structure and their dynamics as a function of size is of interest~\cite{buch04}. A past study shows that the water clusters have played an important role in atmospheric and space chemistry~\cite{Vaida11}, and therefore, cluster studies may contribute to the understanding of the pertinent processes.
    
There have been numerous experimental and theoretical works on various properties and characteristics of water systems. It is the most studied system compared to any other. Debenedetti {\it et al.} have studied the polyamorphism of water and observed supercooled and glassy states well below the freezing point of ice which have been a much-debated topic for the last 30 years~\cite{debenedetti,gomes2019}. A recent study~\cite{canale2019} on unconventional nano-rheology and the high viscosity of melt-water shows two orders of magnitude greater than water in pristine form, leading to slippery ice which is another unique feature of water. To understand these unique phenomena, one needs to go down to the molecular level and study the intermolecular interactions, orientations of atoms, coordination, bonding etc.
    	         
Over the years, studies have been performed at classical scale as well as using {\it ab initio} methods. Geometry optimizations of water clusters have been performed by various methods including density functional theory (DFT) with the Becke’s three-parameter exchange functional along
with the Lee-Yang-Parr correlation functional (B3LYP)~\cite{lenz2005theoretical},  second-order Møller-Plesset perturbation theory (MP2), complete active space self-consistent field (CASSCF), coupled cluster singles and doubles (CCSD) approaches and multiconfigurational complete active space second-order perturbation theory (CASPT2). It has been found~\cite{ip} that CCSD gave the best result of 0.01~\AA~of the average deviation of O-O distances in the water clusters studied as compared to the Hartree-Fock (HF) and CASSCF, which do not involve dynamic electron correlation. In the past, researchers have used a combination of various methods to improve the accuracy of the results. For example, geometry optimization was performed using the MP2 method and the energy calculation was done by 6-311++G(d,p)~\cite{liu2011energetic}, the vertical ionization potential~(VIP) and the adiabatic ionization potential~(AIP) have been studied using  BHandHLYP/6-31++G**, double hybrid B2GP-PLYP and CCSD(T)~\cite{ip} method, whereas the calculation of ionization potential and excitation were done with the help of B3LYP/cc-pVTZ, standard equation of motion coupled cluster singles and doubles (EOM-CCSD) and time-dependent density functional theory (TDDFT)~\cite{ferreira2011electronic}.

Despite the vast literature available on water clusters in the past, there has always been an interest among researchers in order to keep improving their methods to find the best model for water clusters~\cite{wales1998global,james2005global,kabrede2003global,su2004accurate,guimaraes2002global,khan1995examining,defusco2007comparison} such that their calculated values are close to the experimentally obtained results. It is mentioned in the literature that the HCTH functionals~\footnote{The HCTH functional refers to HCTH/407. See reference~\cite{hamprecht1998development,boese2000new,boese2001new,boese2002new}}~turn out to be better than BLYP mainly because of the greater number of parameters that refine the correlation part. The HCTH functional is known to give extremely accurate results for the structures and energies for various systems tested. It was specifically  developed to improve its performance for weak interactions. It contains 15 parameters, which are refined over a training set of 407 molecules, which includes H$_2$O, (H$_2$O)$_2$, and H$_2$O$^+$ molecules. The comparison between the hybrid, meta-GGA and GGA functionals like B97-1, B3LYP, B98, VSXC and HCTH/407, also indicated lowest RMS errors for HCTH/407 with respect to the energies of the anionic and cationic systems, as well as for the ionization potentials and electron affinities. The sum of gradient errors, the shift of the H-bond lengths for H-bonded dimers also show low RMS errors~\cite{hamprecht1998development,boese2000new,boese2001new,boese2002new}.

Thus, as the HCTH functional shows promising results, we have used this functional to test its validity, where we confirmed the structural and electronic properties of water clusters. We found the bond length, bond angles values for monomer and dimeric systems were close to those measured experimentally~\cite{fang2015accurate,chen17,curtiss1979studies,hayashi2000complete}.
For example, for monomer, we obtained, O-H=0.957 \AA\ and H-O-H=104.586\degree\ whereas the experimental values in the literature are O-H=0.957 \AA\ and H-O-H=104.5\degree. Also, our values are better than the other theoretical values as listed in the references~\cite{fang2015accurate} and \cite{chen17}. In the case of dimer, we obtained d(O-O)=3.01 \AA\ which is close to the experimental value of 2.976 \AA\ and with the theoretical values reported~\cite{fang2015accurate}. The binding energy for dimer that we obtained is -5.307 kcal/mol which is close to the experimental value of -5.4$\pm$0.7 kcal/mol~\cite{curtiss1979studies}. Also, the experimental ionization potential for a water monomer reported in literature is 12.6 eV~\cite{hayashi2000complete}, whereas the value we got is 12.85 eV. 

We have also performed an extensive search for global minima of water clusters, for number of molecules with $n=1-20$, employing \textit{ab initio} density functional theory (DFT). We report the minimum energy structures as compared to many extensive works~\cite{rakshit2019atlas,tsai1993theoretical,sadlej1999theoretical,bulusu2006lowest,xantheas2002development}. In addition, we further extend our study to present various other findings of these global minima, which includes structural properties, energetics, bonding, ionization potentials, fragmentation patterns and an interplay between them. We have also presented their optical and vibrational properties, and hence presented a comprehensive study to understand the trend exhibited by water clusters as they evolve in size, from $n=1$ to 20.
  
The organization of the present paper is as follows. In section~\ref{compdetails}, we will describe the model and technical details of our simulations. In section~\ref{results}, we present and discuss the results. In section~\ref{summary}, we will summarize our results along with some important concluding remarks.
	
\section{Methodology}
\label{compdetails}
We have carried out series of calculations on water clusters with $n=1-20$. To begin with, we have obtained low energy isomers using artificial bee colony algorithm of ABCluster package~\cite{zhang15} with TIP4P force field potential~\cite{jorg83}. After eliminating the structures with similar energies ($\Delta E < 0.01$~eV) from tens of thousands of structures, we found at least 200 distinct structures. Out of these a minimum of 10 lowest energy structures were further optimized using the Gaussian 03 package~\cite{g03} with the HCTH/407 functional and the 6-311++G(d,p) basis set~\cite{karen07}.
            
\begin{figure*}[hbt!]
\includegraphics[scale=1.4]{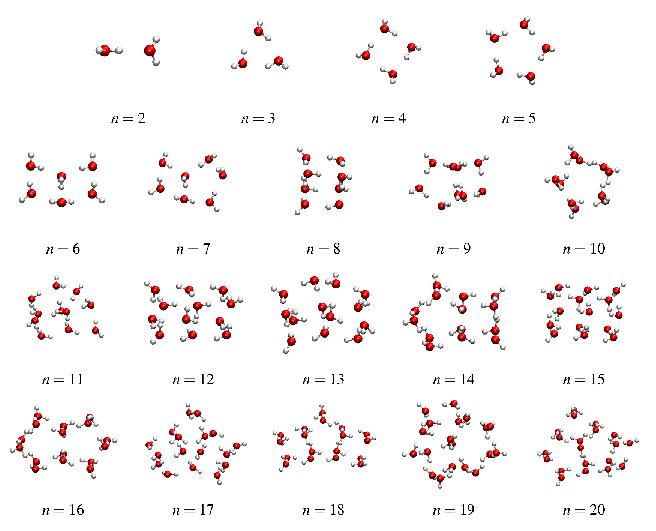}
\caption{Optimized geometries of global minimum water clusters $n=2-20$}
\label{fig:neutral}
\end{figure*}

The optimized coordinates for (H$_2$O)$_n$ clusters, for $n=1-20$ from the Cambridge Cluster Database~(CCD)~\cite{maheshwary2001structure} were compared with those obtained by our calculations. The stacked pentagonal and cage-like structures were found to be lower in energies for $n=11,13-20$ as against the stacked cubic structures available in the CCD. For $n=1-20$, many different conformers were considered out of which the ones we have reported in Figure~\ref{fig:neutral} turn out to be the lowest energy structures. The coordinates of the lowest energy structures are given in the supplementary material.
                
We report here that with the level of exchange correlation used in the present work gives lowest energies as compared with the recent extensive global minima reported by Rakshit \textit{et al.}~\cite{rakshit2019atlas} where they have obtained the putative minima using the TTM2.1-F force field and optimised the structures further at the MP2/aug-cc-pVTZ level. In Table~\ref{tab:comparison} we have also compared our values with the lowest energies obtained by various other groups over the years using different levels of theory and we see that our energy values are the lowest of them ~\cite{tsai1993theoretical,bulusu2006lowest,sadlej1999theoretical,xantheas2002development}. 

\input{table1.tex}

\begin{figure*}[hbt!]
\includegraphics[scale=1.4]{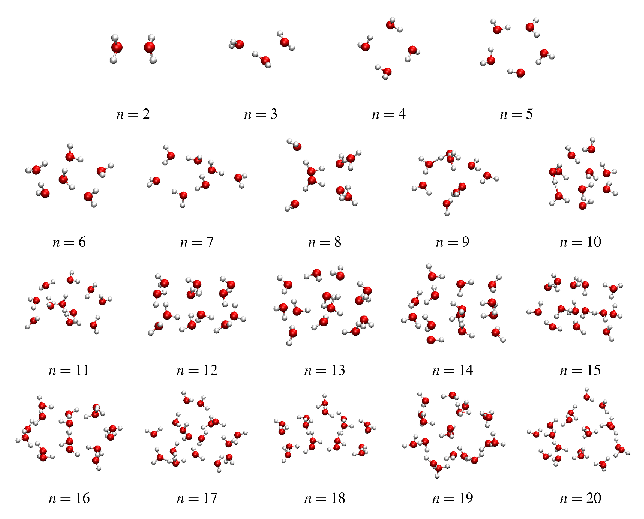}
\caption{Optimized geometries of cationic water clusters for $n=2-20$}
\label{fig:charged}
\end{figure*}

\section{Results and Discussion}
\label{results}
    
\subsection{Structural Properties}

The lowest energy structures of $(\water)_n$ with $n=1-20$ are presented in Figure~\ref{fig:neutral}. The structures for $n=$ 2 to 5 are quasi-planar, linear for $n=2$ and cyclic for $n=3-5$. From $n=6$ we see a significant change in the structural features of these clusters, a transformation into the 3-dimensional structure with respect to the frame of the oxygen atoms is seen. Most of structures can be categorized in to book-shaped, cubic, stacked cubic, stacked pentagonal and cage forms. These structures are preferred over cyclic (or quasi-cyclic) ones as they help maximise the number of H-bonds~\cite{santis2024descriptors} which reduces the electrostatic repulsion between the negatively charged oxygen, and hence are energetically more favourable. Water molecules are stabilized through a hydrogen bond network~\cite{lud2001, maheshwary2001structure}, which was also observed in our work, indicating the vital role it plays in such clusters. The role of cooperative effects that involve all the nearest neighbours has been observed earlier~\cite{znamenskiy2007quantum}.

There have been various studies on water hexamer, with a varied low energy conformers obtained, depending on the level of theory and the basis set used, e.g. Mhin \textit{et al.}~\cite{mhin1994global} obtained that though low-lying energy hexamers are almost isoenergetic, among them, cyclic hexamer is highly stabilized, while Kim \textit{et al.}~\cite{kim1998structures} have found that the lowest energy structure is cage followed by book ($\sim$within 0.1 kcal/mol), however for temperatures above $\sim$40 K, book might be more stable than cage. With our chosen level of theory the first 3-dimensional framework of oxygen atoms is seen for the hexamer which has a book-like shape. Interestingly, here the prism-shaped lowest energy structure of hexamer was transformed into a book-shaped form upon geometry optimization~\cite{leszczynski2012handbook}. Upon the addition of a water molecule, the new molecule gets attached near its edge leading to the widening of the distance between the edge molecules, forming a pentagon-like shape as shown in Figure~\ref{fig:neutral}. However, when one more molecule is added to the heptamer, a significant structural transformation into a nearly cubical cage-like structure is seen for the octamer. Further addition of a water molecule distorts the cubical structure. In the decamer cluster, a bilayer pentagonal structure is observed. Thus, as the cluster evolves, symmetrical structures are observed for even-sized clusters such as for $n=8$~(nearly cubical), $n=10$~(stacked pentagons), and $n=12$~(stacked cubical). We also observe that for $n=14$, the structure is composed of squares and pentagons. Interestingly, the odd-sized $n=15$ cluster is the only one with a symmetric structure. It forms a 3-stacked-pentagonal structure. A similar structure can be seen in the work by \cite{kabrede2003global,wales1998global}. Table-S1 in the supplementary material provides more information with regards to structure and symmetry, namely point group, bond lengths, bond angle etc. for each cluster.

The stability of the water clusters was examined by analyzing the binding energy per molecule  ($\mathrm{E_b}/n$), the first and the second derivatives of the total energy as a function of the number of water molecules  $\mathrm{(\Delta E ~and~ \Delta^2 E)}$, hydrogen bond network, the optical gap between the highest occupied molecular orbital~(HOMO) and the lowest unoccupied molecular orbital~(LUMO), VIP, AIP and fragmentation pattern.
      
\begin{figure}[hbt!]
\includegraphics[width=0.48\textwidth]{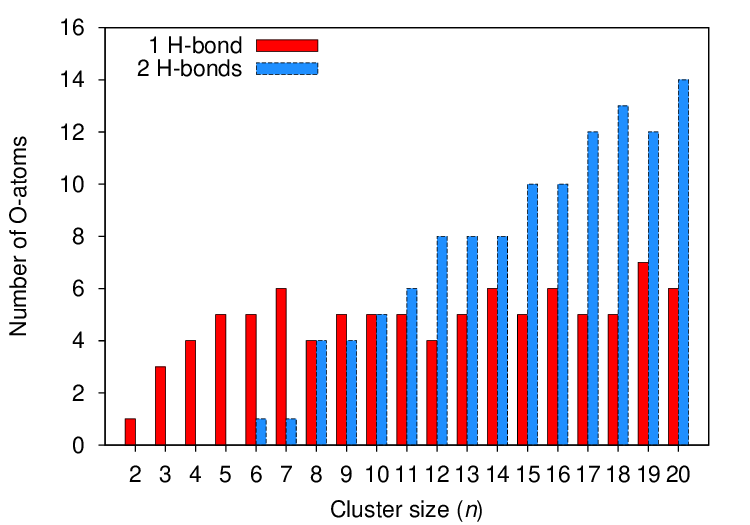}
\caption{Number of hydrogen bonds in water clusters with $n=2-20$}
\label{fig:hbonds_q0}
\end{figure}

\begin{figure}[hbt!]
\includegraphics[width=0.48\textwidth]{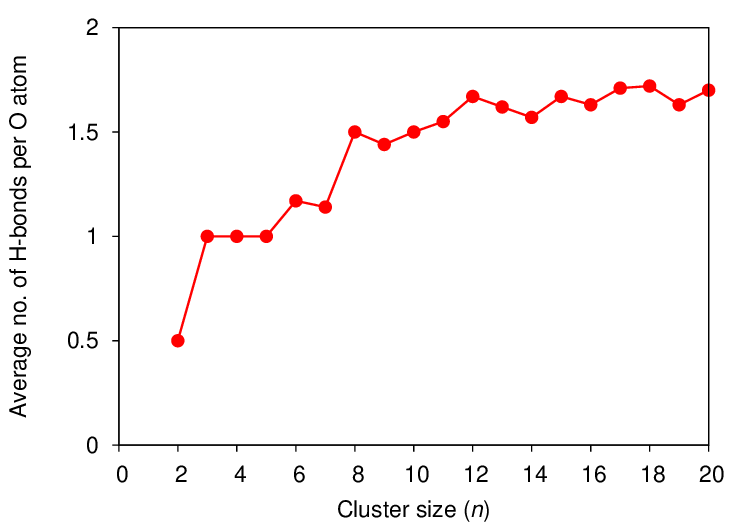}
\caption{Average number of hydrogen bonds per oxygen atom in water clusters}
\label{fig:avg_h_bonds_o}
\end{figure}

To understand the stability of the water clusters, we analyzed the trend in the hydrogen bond network. We plot in Figure~\ref{fig:hbonds_q0}, the number of oxygen atoms having one and two H-bonds. From these figures, it is clear that up to $n=5$, the O atoms only form 1-H bond whereas for $n\ge6$, the oxygen atoms form 1 or 2 H-bonds, giving the clusters more stability as compared to the ring structures, and which is also responsible for the formation of the hydrogen bond network described by Ludwig {\em et al.}~\cite{lud2001}. Maheshwary {\em et al.}~\cite{maheshwary2001structure} have performed extensive studies of different isomers, their stabilization energies, and the role played by the H-bonds using different levels of theories such as Hartree-Fock and DFT. They have found that the average number of hydrogen bonds per water molecule increases with size and gets saturated to about 1.8 when $n\rightarrow20$. We also obtain a similar trend as can be seen in Figure~\ref{fig:avg_h_bonds_o}. The average number of H-bonds per O~atom increases gradually, and levels off between 1.6 to 1.7. On addition of one molecule to $n=19$, the asymmetric structure becomes nearly symmetric now for $n=20$, on getting even no of molecules, with an average no of H-bonds per O atom increasing from 1.63 for $n=19$ to 1.7 for $n=20$, with 14 O-atoms making 2 H-bonds and 6 O-atoms making 1 H-bonds, forming a structure with stacked pentagons and stacked cubic structure.  

\begin{figure*}
\includegraphics[scale=1.3]{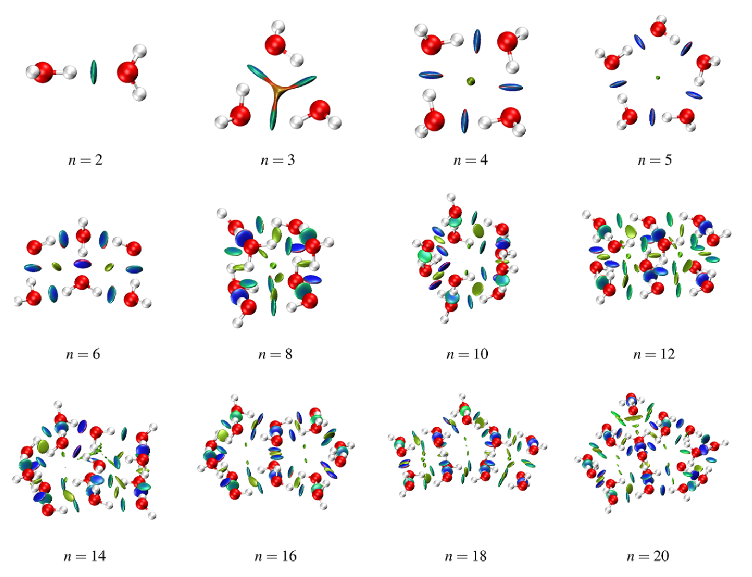}
\caption{Here, various colour shades represent different Non Covalent Interactions (NCI). Shades dark blue to dark green represent strong attraction i.e. H-bonds, the light green shades represent weak Van der Waals interaction, the golden to red shades indicate strong repulsive steric effect in ring and cage structures, and the mix of these colours represent an interplay between these interactions.}
\label{NCI}
\end{figure*}

In addition to this, in order to actually visualize these H-bonds as well as other factors or interactions that contribute to formation of these geometries, we have also performed visual study of non covalent interactions (NCI) using Multiwfn~\cite{lu2012multiwfn} package, and plotted iso-surfaces for strong interactions such as H-bonds between water molecules in clusters, the weak Van der Waals interaction, as well as the strong repulsive forces or the steric effects arising from the closed ring structures. These can be seen in Figure~\ref{NCI}. We have shown here small clusters as well as even number of clusters some of which are symmetric. Thus, in dimer we can see an iso-surface of H-bond. Trimer being the smallest ring structure and the molecules being close to each other, there is a steric hindrance observed at the center of the cluster, along with the H-bonds between the molecules. As we go further, in tetramer we can also see the weak Van der Waals interaction, at the center, between the H-bonds. The number of iso-surfaces of H-bonds and the Van der Waals interaction go on increasing as we go up to $n=20$. Thus, as we can see from these figures, these non covalent interactions, play an important role in binding the cluster together and giving them stability. The H-bond network leads to the formation of clusters of water molecules and giving it various unique properties at nano-scale.

\subsection{Energetics}
\subsubsection{Binding Energy}
\begin{figure}
\includegraphics[width=0.47\textwidth]{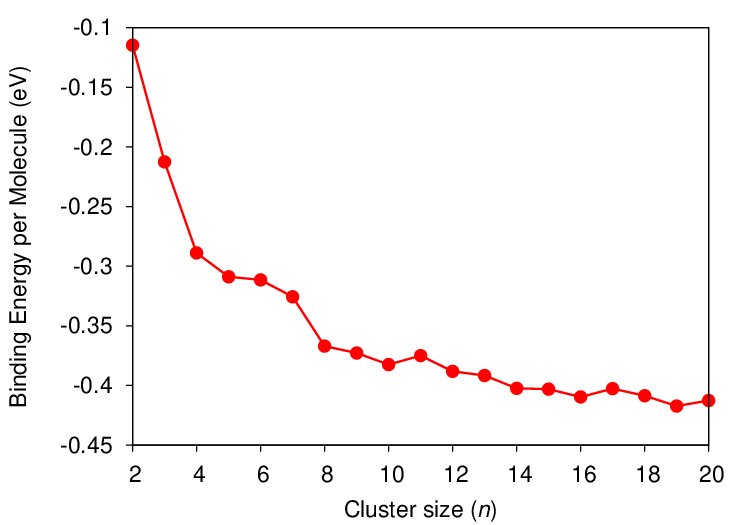}
\caption{Binding energy per molecule as a function of cluster size ($n$)}
\label{fig:ben}
\end{figure}

To see the stability of water clusters, we have computed the binding energy per molecule in water clusters as a function of cluster size which indicates how easy or difficult it is for the water molecules to form a cluster. It is calculated as follows:
$\mathrm{E_b\textbf{/$n$} = \frac{1}{\textit{n}}\Big(E(\water)_\textit{n}-\textit{n}E(\water)\Big)}$, 
where $\mathrm{E(\water)_\textit{n}}$
is the total interaction energy of the $n$-mer water cluster and E$(\water)$ is the energy of the water monomer. This can be seen in Figure~\ref{fig:ben}.

The experimental value of Binding Energy for dimer has been found at $-5.4\pm0.7$ kcal/mol~\cite{curtiss1979studies}, whereas in the present study we have obtained it to be at $-5.307$ kcal/mol which is in good agreement. Table-S2 in the supplementary material shows a comparison of binding energies in this work with the results of other researches using various techniques. The energy difference obtained is due to different approximation methods used by each of them. Our method gives binding energies that are higher than those obtained in reference~\cite{lenz2005theoretical} (except for $n=$ 11, 12 and 20), employing B3LYP, aug-cc-pVDZ basis set.

It is noted that the increased stability of the clusters as the size increases and approaches the bulk, where it saturates, is in the best agreement with the reported results~\cite{ade04}.
However, from  Figure~\ref{fig:ben}, we see a greater increase in the $\mathrm{E_b}/n$ for $n=4$ because of its square shaped symmetry, and a further notable increase in $\mathrm{E_b}/n$ for $n=8$ due to its cubical geometry. As we go further, we see that for even number of clusters, i.e. for $n=$ 10, 12, 14, 16, 18, the $\mathrm{E_b}/n$ values are slightly greater than their preceding odd numbered motifs.
        
\begin{figure}
\includegraphics[width=0.48\textwidth]{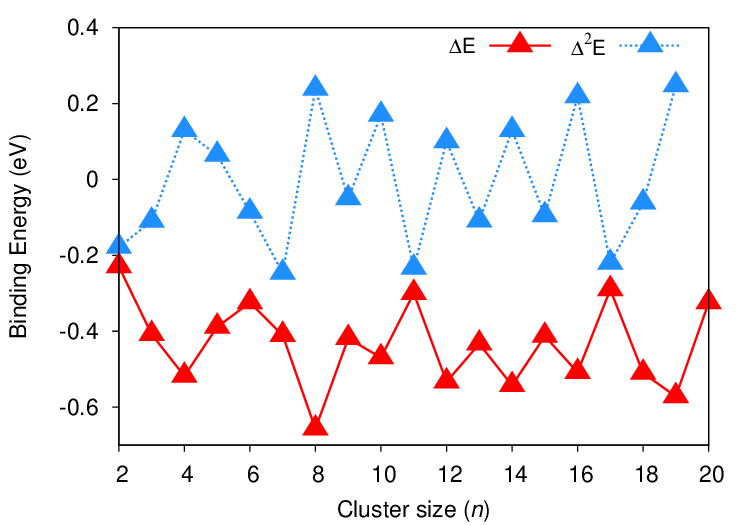}
\caption{Binding energy and $\Delta^2$E}
\label{fig:den}
\end{figure}

To support the above arguments and verify the predicted stability of the clusters, we have also calculated binding energies in the form of their first and second energy differences i.e. the energy required to remove one molecule from a cluster and its second order derivative,  $\mathrm{\Delta{E}}$ $\mathrm{( \Delta E = E_\textit{n} - ( E_{\textit{n}-1} + E_1 ) )}$~and~  $\mathrm{\Delta^2E}$ $\mathrm{(\Delta^2E = E_{\textit{n}+1} - 2E_\textit{n} + E_{\textit{n}-1})}$ respectively for all water clusters. It is shown in Figure~\ref{fig:den}. A low value of $\Delta{E}$ along with a large value of $\Delta^2 E$ indicates a stable cluster. Structures and symmetry of the clusters as well as the number of H-bonds that each O atom in the cluster makes, play a major role in determining their binding energies. We see that compared to odd number of clusters, even number of clusters have higher binding energies, except hexamer, with a lower binding energy, due to its book shaped structure which is less stable as an electron can be removed from its edges with less number of co-ordination of H-bonds it makes, i.e. only one O atom makes 2 H-bonds, rest all O-atoms make one H-bonds. Higher binding energy implies higher stability. Octamer, shows higher binding energy with its cubical shape (-ve sign represents the bound state of molecules in the cluster). Similarly tetramer with square shape, decamer with stacked pentagons, cluster with $n=12$ with stacked cubic, $n=14$ with cubic and stacked pentagons, $n=16$ with stacked pentagons, and $n=19$ with cage structure show high binding energies. Whereas, cluster with $n=11$ formed by adding one molecule to the decamer can easily be removed and cluster with $n=17$ also being an asymmetrical structure, removal of electron does not need much of energy. Thus $n=$ 11 and 17 clusters show low binding energy and thus lowest cluster stabilities.
        
Thus, to summarize, from Figure~\ref{fig:ben} and ~\ref{fig:den}, we can observe that all the clusters with an even number of water molecules seem to be more stable than the odd-numbered, except for $n=$ 2 and 6. The clusters with $n=$ 4, 8, 12, 14, 16 and 19 water molecules are found to be more stable as compared to others. From the literature it can be seen that the cuboid or the fused pentamers or their combination show a greater stability~\cite{maheshwary2001structure} or the conformers with four membered rings, namely $n=$ 4, 8, 12 are relatively more stable than the other clusters~\cite{lee1995chemical}. The central difference approximation has also been used~\cite{kazachenko2009improved} to illustrate that even number of clusters are more stable than the odd numbered, e.g. $n=$ 8, 12, 16, 20. Thus, we can see that the structure, symmetry of the clusters and the number of H-bonds per O-atom, play a significant role in determining their stability.

\subsubsection{Shape Analysis}
\begin{figure}[hbt!]
\includegraphics[width=0.48\textwidth]{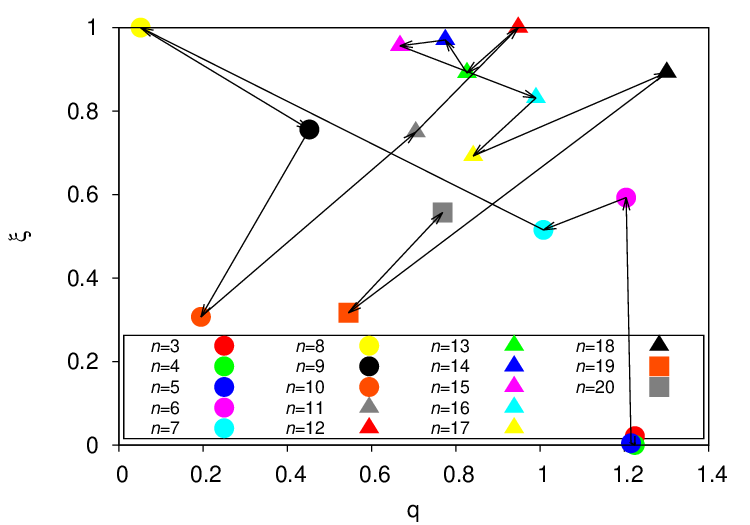}
\caption{Shape parameter as a function of deformation of the size of the cluster.}
\label{fig:shape}
\end{figure}

We have analyzed the shape of the water clusters using the quadrupole deformation parameter  $\epsilon_{d}=\frac{2Q_x}{Q_y+Q_z}$, where  $Q_x~\ge~Q_y~\ge~Q_z$ are the eigenvalues of the quadrupole tensor $Q_{ij}= \sum_{I} R_{Ii}R_{Ij}$. Here $R_{Ii}$ is the $i^{th}$ coordinate of ion $I$ relative to the centre of mass. When all the eigenvalues are equal, $i.e.~Q_x=Q_y=Q_z$, $\epsilon_{d}=1$, Such a system has perfect spherical symmetry. Any deviation of $\epsilon_{d}$ from this value indicates quadrupole deformation of some sort. Further insights can be obtained by analyzing the dimensionless Hill-Wheeler parameters~\cite{bohr1998nuclear}. Whether the cluster is oblate or prolate can be determined using the parameters $\xi=\frac{\overline{Q_x}-\overline{Q_y}}{\overline{Q_x}-\overline{Q_z}}$, where $0\leq\xi\leq1$. The value $\xi=0$ corresponds to an oblate structure, whereas $\xi=1$ corresponds to a prolate structure. Any value of $\xi$ between 0 and 1 implies a triaxial deformation. A drawback with this approach is that for spherical structures, $\xi$ becomes indeterminate and thus can lie anywhere between 0 and 1, and hence it becomes difficult to identify. This can be overcome by analyzing $\xi$ against another parameter $q$, given by $q^2=\frac{\overline{Q_x}^2+\overline{Q_y}^2+\overline{Q_z}^2}{Tr(Q)^2}$. The parameter $q$ determines the measure of sphericity of the system with $q=0$, indicating spherical shape, and as it goes on increasing the structures tend to become less spherical, $i.e.$ having a greater deformation. In Figure~\ref{fig:shape}, we have plotted the values of $\xi$ against $q$  for all the clusters. One can observe that the structures with $\xi$ between 1 and 0.5, namely $n$ = 13, 14, 15, 16 and 18 have prolate character, whereas those with $\xi$ between 0 and 0.5, $i.e.~n$ = 10, and 19 have an oblate character.
        
Analyzing the contribution of the eigenvalues $Q_x,~Q_y$ and $Q_z$ (refer to Figure-S1 in supplementary material), we can infer that the clusters of sizes $n=$~3, 4 and 5 with $\xi=0$ are oblate, whereas $n=$~8 and 12 with $\xi=1$ are prolate. For $n=8$, with a $q$ value of $0.05$, indicates a nearly spherical symmetry with a slight distortion of an oblate nature. A similar pattern is also followed by the stacked pentagonal decamer, the nonamer structure, and the cage structure of the $n=19$ cluster, albeit with slightly higher distortions. It is interesting to note that most of the clusters whose $\xi$ and $q$ values lie in the central region ($i. e.~n=$ 6, 7, 9, 11, 17 and 20 show a significant distortion in their structures and are some of the least stable clusters as noted above from figure~\ref{fig:den}. The 2-dimensional clusters with $n$ = 3, 4 and 5 and the first 3-dimensional structure of $n=6$ exhibit very high values of the $q$ parameter indicating a substantial deviation from spherical symmetry. The highest value of $q=1.3$ was found for $n=18$ which has maximum non-spherical distortions.

\subsubsection{Fragmentation}
               
\begin{figure}
\includegraphics[width=0.48\textwidth]{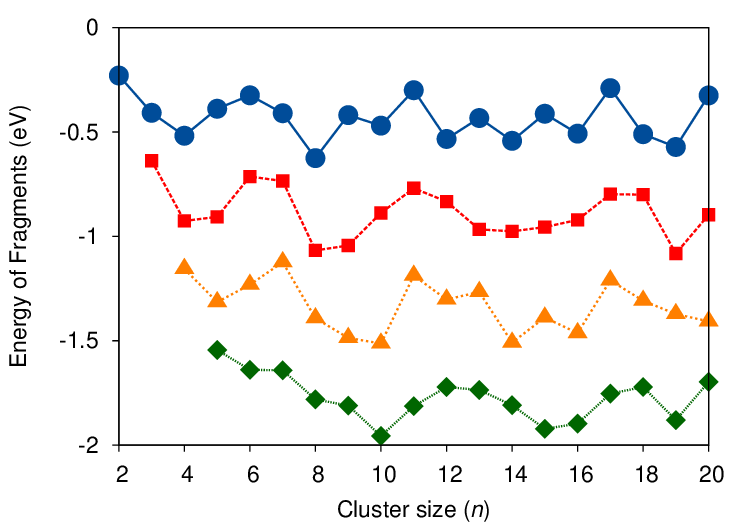}
\caption{Fragmentation of water clusters $n=$ 2 - 20 into 2 (Blue), 3 (Red), 4 (Yellow), 5 (Green) fragments respectively}
\label{fig:fragments}
\end{figure}

Analyzing the fragmentation patterns of water clusters offers valuable insights into their stability.
We have calculated the fragmentation energies for all clusters, considering all possible breakups from 2 to $n$-fragments, where $n$ is the number of molecules in the cluster. In figure~\ref{fig:fragments}, we present the lowest-energy fragmentation pathways for all clusters, specifically, the breakup into 2, 3, 4 and 5 fragments. The fragmentation energy for $m$-fragments is calculated using the following:
$$ \mathrm{
   \Delta E_{\textit{m}}^{\textit{l}_1,\textit{l}_2,\cdots,\textit{l}_\textit{m}} = E_\textit{n} - \sum_{\textit{i}=1}^{\textit{m}} E_{\textit{l}_\textit{i}} 
    } $$
where, $\mathrm{\sum_{\textit{i}=1}^\textit{m}}$ $l_i = n$ with $m<n$, $l_i$ being the number of molecules in the $i^{th}$ fragment. The alphabets $n$ and $m$ represent cluster size and number of fragments, respectively. E$_n$ is the total energy of the cluster and E$_{l_i}~(i=1-m)$ are the energies of the fragments. A large fragmentation energy indicates a strong bonding between the water molecules which in turn makes that cluster more stable. From the figure, we see that for 2 fragments the energy needed to break the clusters is the lowest. As the number of fragments is increased, the fragmentation energy goes on increasing. This is consistent with the experimental results, where the dissociation energies of the neutral water clusters in the size range $n=2-9$ suddenly increase from dimer to trimer, followed by the gradual change in energy with an increase in cluster size~\cite{belau2007vacuum}. The figure also shows a dip in the fragmentation energies for certain fragment sizes with $m=$ 4, 8, 10, 16 and 19. Incidentally, these clusters were found to be some of the most stable ones (see figure~\ref{fig:den}).

\begin{figure}[hbt!]
\includegraphics[width=0.48\textwidth]{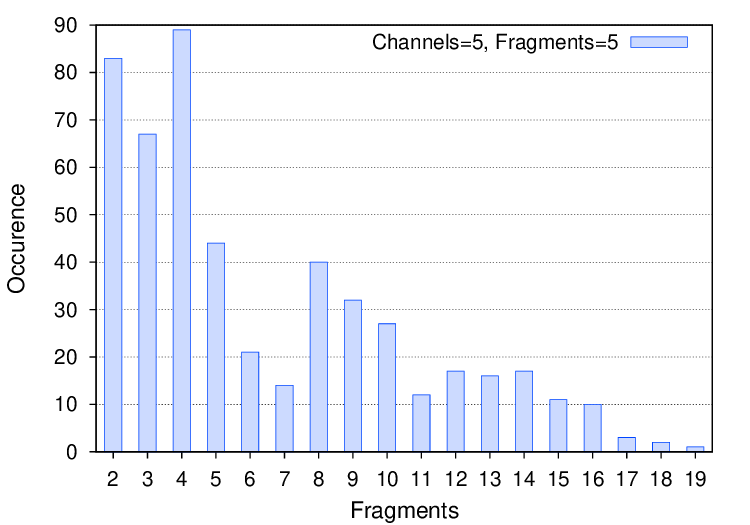}
\caption{Frequency of occurrence of different sizes of fragments during fragmentation process for the lowest five channels.}
\label{fig:occ-fragments}
\end{figure}

In figure~\ref{fig:occ-fragments}, we show the number of occurrences of all the fragments found in the lowest five channels. The frequency of water monomers is not displayed, as they are the predominant fragment in all channels and represent the most favourable pathway due to their low dissociation energy.  The formation of monomers is observed in a multi-step fragmentation process with the loss of a single water molecule by every water cluster. This is followed by the maximum occurrence of the dimer, which is a second-step fragmentation process. As the fragmentation channel increases, the likelihood of large fragment formation decreases, impeded by a substantial potential barrier, when all possible fragmentation pathways are taken into account. However, we also note another interesting feature. Despite the declining trend in the histogram, the occurrence of the fragments $m=$~4, 8, 12 and 14 show higher values than their neighbouring fragments, implying that these fragments are more likely to be observed in fragmentation experiments. These results are consistent with the reported data~\cite{liu2011energetic}, where the stability of clusters has been explained with the first and second-best fragmentation channels. Thus, a higher fragmentation energy of any fragment indicates the fragment to be more stable. Thus, the fragmentation process shows an increase in the stability with the size of the cluster which supports the observation of binding energy per atom as described above.

\subsection{Electronic Properties}
\begin{figure}[hbt!]    	
\includegraphics[width=0.48\textwidth]{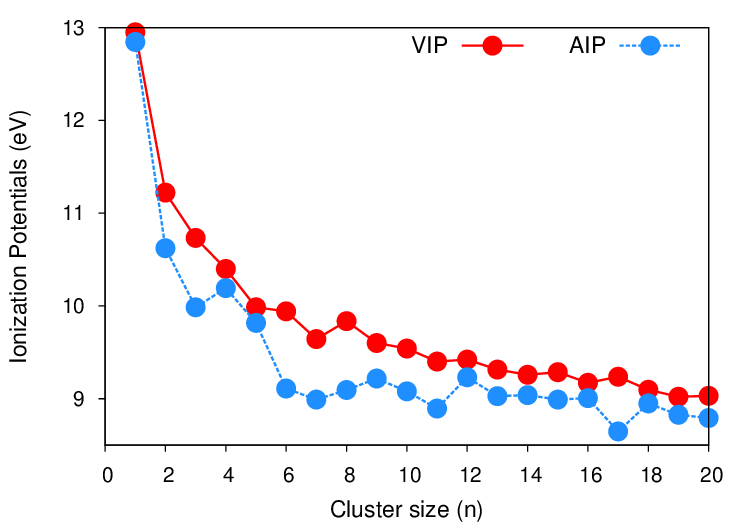}
\caption{Vertical and Adiabatic ionization potentials of water clusters for $n=1-20$}
\label{fig:ip}
\end{figure}

The stability of the  $(\water)_n$ clusters is also explained using ionization potential. The energies of neutral and singly charged cationic water clusters were calculated with and without the geometry optimization. Using the energy values from these we calculate the VIP and the AIP respectively.	In experiments, it is difficult to measure VIP since such clusters almost undergo instantaneous structural relaxation upon removal of an electron, making the AIP more important for us to study. Figure~\ref{fig:ip} shows the VIP~(red curve) and the AIP~(blue	curve). Since for the monomer, it gets difficult to remove an electron from its filled molecular orbital, the value of its ionization potential~(IP) is the highest, and goes on decreasing with size. It is to be noted that the AIPs are lower than the VIPs. However, for certain sizes such as $n=2,3,6,7,8,11$ and 17, the difference is much larger. This can be attributed to the significant change in their geometric structures after the removal of an electron~(see Figures~\ref{fig:neutral} and~\ref{fig:charged}). For instance, the two molecules in the neutral dimeric cluster are bonded by a single H-bond, whereas in the positively charged form, after structural relaxation, a change in orientation of molecules lead to two H-bonds - one for each O-atom, leading to an increased difference between the VIP and AIP. The neutral trimer cluster forms an equilateral triangle geometry which changes to a scalene triangle upon geometric relaxation after the removal of an electron. On the other hand, the changes in the structure for clusters with sizes $n=4$ and 5 are insignificant.  The experimentally measured~\cite{belau2007vacuum} appearance energies of neutral water clusters for $n=2-80$ have been found to converge to around 10.6~eV for $n>20$, IP for a single water molecule is 12.6 eV~\cite{hayashi2000complete}, whereas the AIP for liquid water has been reported to be at 9.3 eV~\cite{ip}. Incidentally, the ionization energies, \textit{i.e.} VIP and AIP in our work starts from about 12.95 eV and 12.85 eV for $n=1$ and reaches a value of about 9.03 eV and 8.79 eV respectively for $n=20$.

\begin{figure}[hbt!]
\includegraphics[width=0.48\textwidth]{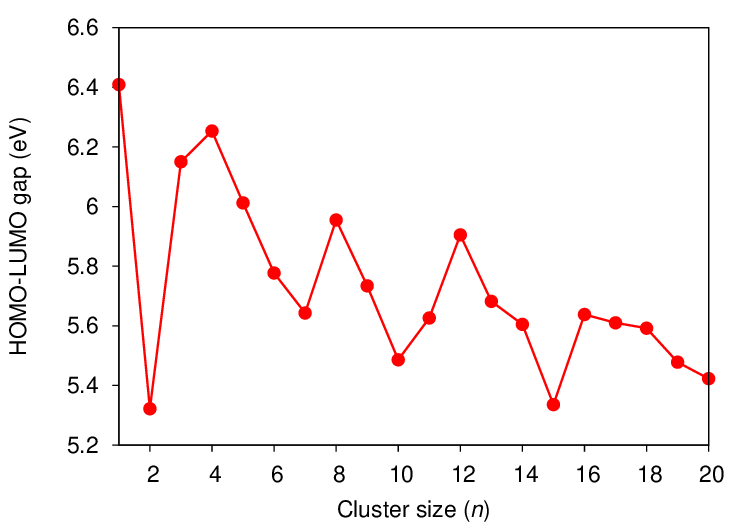}
\caption{HOMO-LUMO gap of water clusters}
\label{HOMO-LUMO gap}
\end{figure}

The quantification of eigenvalue spectra allows us to understand the factors that affect the stability of dynamical systems. We have examined the eigenvalue spectrum as a function of cluster size (See Figure-S2 in supplementary material). The red and blue energy levels are identified as the occupied and unoccupied orbitals respectively. From the HOMO-LUMO gap, we can understand the stability of the structure. We see that for dimer it is smaller than that of monomer, indicating that it is less stable. From Figure~\ref{HOMO-LUMO gap} it can be seen that $n=4,8,12,16$ show larger HOMO-LUMO gaps, indicating that they are chemically more stable than the other clusters. 
        
\subsection{IR Spectra}
\begin{figure}[ht!]
\includegraphics[width=0.48\textwidth]{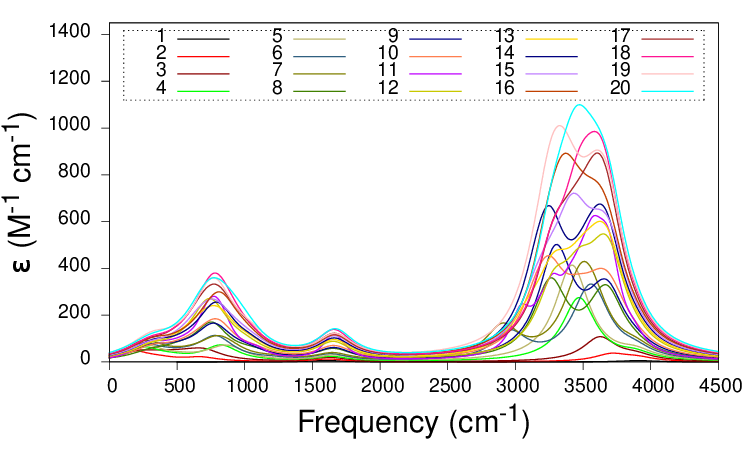}\\
\caption{Infrared spectra of water clusters for $n=1-20$}
\label{fig:ir_raman}
\end{figure}

Vibrational spectroscopy deals with the infrared spectra of the molecules which are specially used for the elucidation of molecular structure. The infrared spectra of the ground state geometries of water clusters with $n=1-20$ are shown in Figure~\ref{fig:ir_raman}. First, we begin with water monomer followed by clusters with $n=2,4,8,12$ and 16 which are found to be the most stable structures from the energetic, fragmentation and HOMO-LUMO gap analysis.

In the case of monomer, we observed three IR bands with vibrational modes of bending, symmetric and  asymmetric stretching. The frequency of the symmetric and asymmetric stretching modes are 3813.12~cm$^{-1}$ and 3918.87~cm$^{-1}$ respectively whereas the bending mode is observed to be at 1602.44~cm$^{-1}$. It is also close to the reported experimental and theoretical results~\cite{otto14, kim94, devl01}. 
                
In the case of the water dimer, we have observed vibrational and rotational modes. The rotational modes have been observed at a lower frequency range as compared to the vibrational modes. Six rotational modes are lying in the range (141-675~cm$^{-1}$). The four bands at 3705.57~cm$^{-1}$ (symmetric), 3814.40~cm$^{-1}$ (symmetric), 3892.69~cm$^{-1}$ (asymmetric) and 3915~cm$^{-1}$ (asymmetric) in the  $(\water)_2$ geometry are ascribed to stretching vibrations and the bands at 1613.15~cm$^{-1}$ and 1629.22~cm$^{-1}$ are assigned to the bending vibrations of O-H bonds. It is found that our results are in fairly good agreement with those obtained from the theoretical and experimental results~\cite{ben56, kim94, devl01,10.1063/1.4936654}. It is observed that the bonded H-O-H bending mode of vibration is blue-shifted compared to the water monomer's H-O-H bend. On the other hand, the O-H symmetric and asymmetric stretching modes show red shifts In the case of water trimer, we have observed H-O-H bending modes at 1624.24~cm$^{-1}$, 1629.86~cm$^{-1}$ and 1651.96~cm$^{-1}$, and six O-H stretching modes between 3560-3894~cm$^{-1}$. The observed values of the frequencies show blue- and red-shift in the H-O-H bending  and O-H stretching respectively and are consistent with the reported results~\cite{anamika18,10.1063/1.4936654,Howard}.

The H-O-H bending and O-H stretching modes in cyclic water tetramer show a very interesting pattern due to the arrangement of the equivalent hydrogen bonds on the sides of the square. Each O-H group directly couples with its hydrogen bond acceptor partner. The four H-O-H bending and eight O-H stretching modes belong to 1630-1681~cm$^{-1}$ and 3376-3886~cm$^{-1}$, respectively, which shows the blue and red-shift for the water monomer, dimer and trimer. All the calculated values of vibrational frequencies are in reasonable agreement with the reported result~\cite{otto14,Howard}. Thus, this indicates that increasing the cluster size, which brings in the role of hydrogen bonds, has a substantial effect on the vibrational spectra of water molecules. Two modes of O-H stretching in each case along opposite corners of the square, sides of the square, and inside and outside the plane were observed.

Next, in stable water octamer the IR vibrational spectrum shows that the eight H-O-H bending modes lie in the range 1634-1728~cm$^{-1}$. On the other hand, the sixteen stretching modes belong to 3204-3878~cm$^{-1}$. The calculated values of the vibrational frequency of the stretching modes are in excellent agreement with the reported results~\cite{christoper98}. The detailed analysis of the vibrational frequency of the O-H stretching mode is classified into three regions which are associated with free O-H, single donor O-H and double donor O-H groups. It is seen that four free O-H vibrations are nearly degenerate (3878.32~cm$^{-1}$, 3878.34~cm$^{-1}$, 3878.41~cm$^{-1}$ and 3878.73~cm$^{-1}$). The four single-donor O-H stretch modes, with their strong, linear H-bonds,
have large frequency shifts of 186~cm$^{-1}$ to 263~cm$^{-1}$. On the other hand, the double donor O-H has a higher frequency shift of 434-488~cm$^{-1}$ with respect to single donor O-H. It is also observed that there are two nearly degenerate modes in each case (single donor O-H and double donor O-H). It is the highest-frequency single-donor mode, which is nondegenerate (3630~cm$^{-1}$ and 3648~cm$^{-1}$). It is also observed that the double-donor vibrations are weaker than the single-donor O-H stretch vibrations because of the weaker, nonlinear H-bonds created when both O-H groups on a given water molecule are involved in H-bonds in a strained structure of water octamer. This results in a gap of 411~cm$^{-1}$ between single and double donor fundamental frequencies. It is found that the four highest and lowest frequency modes have asymmetric and symmetric O-H stretch character on double donor water molecules. It is observed that the calculated values of the vibrational frequencies of bending and stretching modes are comparable to the reported theoretical values~\cite{hubert07}.

Now we look at the IR vibrational spectrum of (H$_2$O)$_{12}$. The vibrational spectrum shows three distinct regions namely, intermolecular vibrations O$\cdots$H between 59~cm$^{-1}$ and 1095~cm$^{-1}$, a region of intramolecular H-O-H bending modes between 1639~cm$^{-1}$ and 1735~cm$^{-1}$, and a region of intramolecular O-H stretching vibrations between 3241~cm$^{-1}$ and 3877~cm$^{-1}$. All these regions show doubly degenerate modes. The vibrational spectrum of (H$_2$O)$_{16}$ also shows a similar trend, and in addition to it, there is a red-shift of intramolecular bands for the frequency range from 3241~cm$^{-1}$ to 3189~cm$^{-1}$. Previous studies~\cite{seki2020bending,falk1984frequency} have found that due to the H-bonds formation, the frequency of the H-O-H bending mode increases from gas to liquid phase. As we have seen in our results, we find a similar trend of increase in the H-O-H frequency as the cluster size goes on increasing. It is also stated that, a greater H-O-H bending mode frequency indicates a lower O-H stretching frequency, implying stronger intermolecular H-bonds. In the present work as we have seen blue-shift of H-O-H vibrational mode along with red-shift of O-H stretching mode for $n=2, 3, 4, 8, ..$, this confirms an increase in the H-bond strength with increasing $n$.
            
\subsection{Optical Properties}

\begin{figure*}
\begin{tabular}{ll}
\includegraphics[scale=0.38]{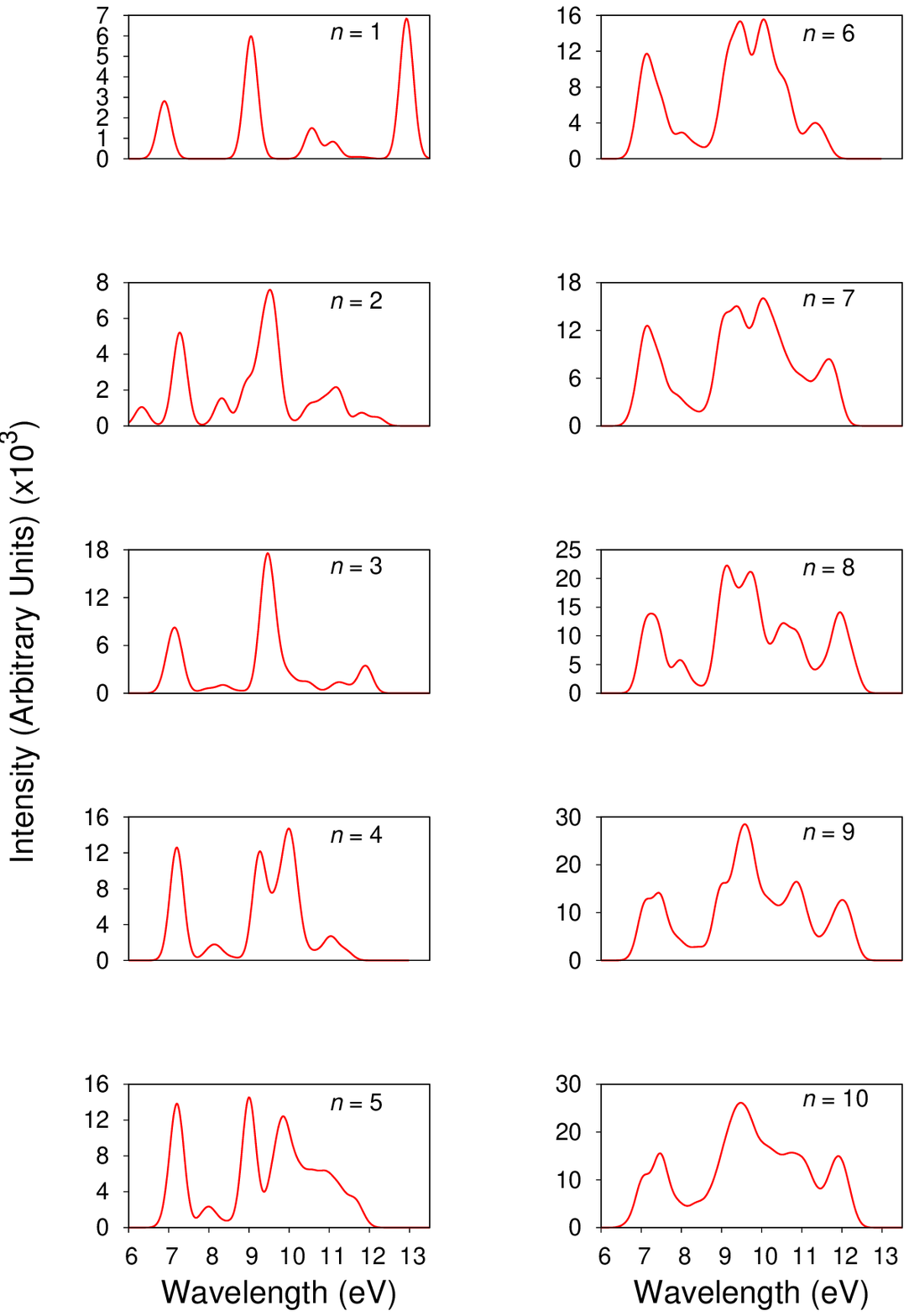}&
\includegraphics[scale=0.38]{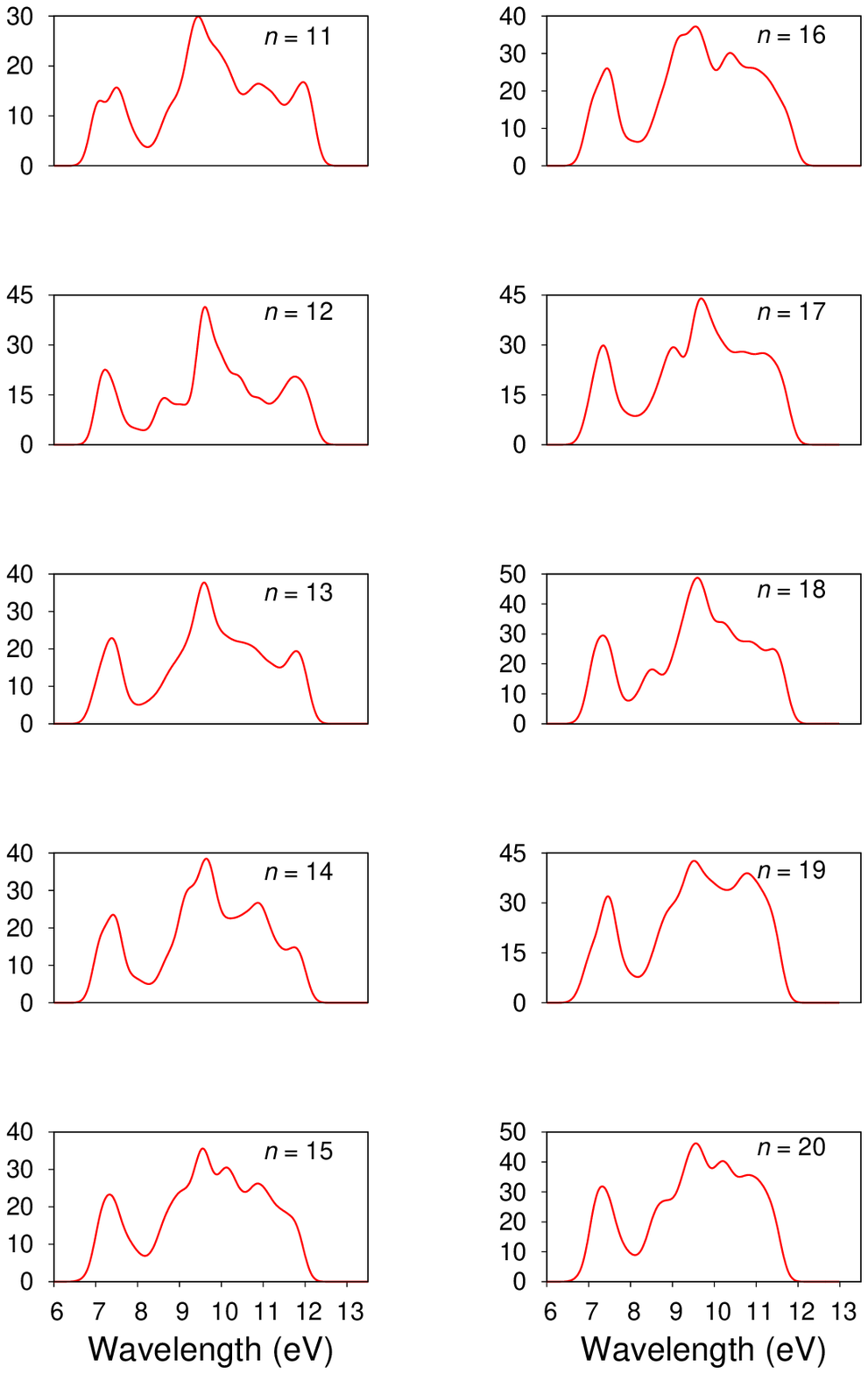}\\
\end{tabular}
\caption{Optical absorption spectra of water clusters for $n=1-20$}
\label{fig:optics}
\end{figure*}

To understand the electronic transitions and its origin, we have carried out TDDFT calculations on the optimized water clusters for $n=1-20$, using B3LYP functional~\cite{bauernschmitt1996treatment} and 6311++g(d,p) basis set. The n-states chosen for each cluster were such that they covered all the states upto the first ionization potential.
The origin of optical excitations in these clusters is explained with the help of the eigenvalue spectra, optical bandgap and molecular orbitals (MOs). Table-S3 given in the supplementary material represents the summary of the electronic states of the first three major excitations of water clusters with high oscillator strengths. It shows the VIP and AIP, excitation energy of major transitions  $\mathrm{(E_{exc})}$, energy of the most significant peak  $\mathrm{(E_m)}$, oscillator strength $\mathrm{(f_{osc})}$, optical bandgap $\mathrm{(E_{gopt})}$ and excited state compositions, their assignments, along with their percentage contribution. The symbols `H' and `L' stand for the HOMO and the LUMO respectively, and corresponding states below (H$_1$, H$_2$, ...) and above (L$_1$, L$_2$, ...) these levels. Figure~\ref{fig:optics} shows the optical spectra of all the clusters. 

A number of features can be discerned from the table and the 
optical spectra. 
i. The optical bandgap varies from 7.14 eV to 8.17 eV which is oscillatory in nature with maximum value at $n=4,8,12,16$ 
and belongs to ultra-violet region.
ii. As the cluster size increases for n$\ge$10, the number of spectral lines increases along with line broadening. 
iii. We find that the optical spectra of the stable structures show blue-shifts. This can be attributed to the stability of the clusters. 
iv. It is also observed that only the stable structures show maximum oscillator strength.
v. It is interesting to note that in all the cases the first excitation occurs between HOMO to LUMO. We will now explain and discuss each feature in the next section.

\begin{figure}[hbt!]
\includegraphics[width=0.48\textwidth]{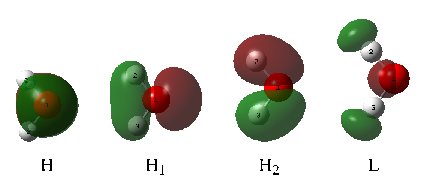}
\caption{MOs involved in first three excitations for $n=1$, with percentage contribution $\ge50\%$}
\label{MOs_1}
\end{figure}

We begin our discussion with water molecule. The absorption spectrum of a fully structurally relaxed monomer can be seen in Figure~\ref{fig:optics}. It shows the three major peaks at 6.89~eV, 9.05~eV and 12.93~eV. The optical bandgap obtained is 8.17~eV, which is within 10\% of the experimental value~\cite{hermann2008resolving}. From Figure~\ref{MOs_1}, it is observed that the first excitation takes place between non-bonding HOMO, which is $\pi_y$ to anti-bonding LUMO, that is $\sigma_z^*$. The HOMO to LUMO transition is obtained at 6.89 eV (179.88 nm) with oscillator strength of 0.04. 
The second major transition occurs between H$_1$ to L, $i.e.~\sigma_x$ (bonding) to $\sigma_z^*$ (anti-bonding) orbital with oscillator strength of 0.09 at 137.01 nm. The third major transition from H$_2$ (bonding orbital $\sigma_z$) to L (anti-bonding $\sigma_z^*$) involves still higher energy, thus, a further blue-shift is seen, with the highest intensity obtained at 12.93 eV (95.92 nm), with an oscillator strength of 0.10. The other probable electron transitions with their respective oscillator strengths give us different peaks in the UV-VIS spectrum with energy corresponding to the energy difference between those two levels. We can see these energy levels from the eigenvalue spectra.

\begin{figure}[hbt!]
\includegraphics[width=0.38\textwidth]{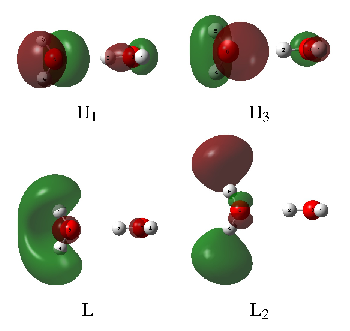}
\caption{MOs involved in first three excitations for $n=2$, with percentage contribution $\ge50\%$}
\label{MOs_2}
\end{figure}

When two water molecules come together to form a dimer, there is a splitting of energy levels. The HOMO for the dimer is at a higher energy and the LUMO is at a lower energy than that of the monomer. Thus reducing the bandgap as compared to that of monomer and thus lesser energy is needed for electron transition from HOMO to LUMO, leading to a red shift in the bandgap. In fact, for the dimer, the optical bandgap obtained is the lowest among all the twenty clusters. We can also see an overall red-shift in its optical spectrum. The orbitals involved for the most probable transition, with $\mathrm{f_{osc}}=0.09$ are mainly contributed by the H$_3$ and L, having bonding and anti-bonding character respectively as shown in Figure~\ref{MOs_2}. These orbitals are localized over individual molecules, with H$_3$ exhibiting $\sigma_x$ type and L showing $\sigma_z^*$ type characteristics. It is the tenth excitation and needs an energy of 9.6 eV (129.2 nm). The second major peak at 7.3 eV is due to excitation from H$_1\rightarrow$~L, having $\sigma_x$ type (bonding) and $\sigma_z^*$ type (anti-bonding) attributes respectively, with oscillator strength of 0.06. The third major peak with $\mathrm{f_{osc}}=0.02$ is seen at 11.23 eV due to $\sigma_x$ type H$_3$ (bonding orbitals) to ~L$_2$ (anti-bonding orbitals) transition. 

Going further on to trimer, we find that for $\mathrm{f_{osc}}=0.12$, the orbitals involving the electronic transition are H$_5$~(having bonding) and L~(having anti-bonding character). But it is seen that H$_4$ and H$_5$ are nearly degenerate. Because of its equilateral triangle shape or the smallest ring shape, there is an overlap of its $\sigma_x$ type orbitals over two of its molecules in the H$_5$ state. This gives it stability, requiring greater energy for electron removal and thus a larger bandgap is observed. Other major contribution for this excitation comes from $\sigma_x$ type H$_3$~(bonding orbitals) and L$_1$~(anti-bonding orbitals).

\begin{figure}[hbt!]
\includegraphics[width=0.44\textwidth]{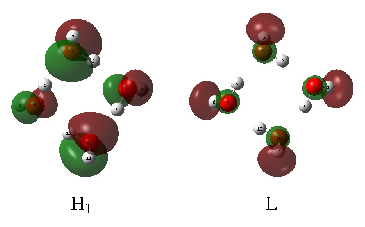}
\caption{MOs involved in first three excitations for $n=4$, with percentage contribution $\ge50\%$}
\label{MOs_4}
\end{figure}

For tetramer, the HOMO belongs to a set of four degenerate orbitals- H to H$_3$ (all having non-bonding attributes), which again is the reason for greater stability of the cluster, thus needing higher energy for excitation. This can also be seen from the percentage contribution of MOs involved in the highest probable transition having $\mathrm{f_{osc}}=0.1$ at 9.98 eV, with major contributions coming from H~$\rightarrow$~L$_5$, H$_1$~$\rightarrow$~L$_6$, H$_2$~$\rightarrow$~L$_7$ with 34.56$\%$, 26.49$\%$, 23.54$\%$ respectively. The second major peak is due to the excitation of bonding H$_6$ to anti-bonding L with $\mathrm{f_{osc}}=0.09$. The third highest peak comes from the transition H$_1$~$\rightarrow$~L (non-bonding to anti-bonding orbitals respectively) with oscillator strength of 0.07. Here, H, H$_1$, H$_2$ have $\pi_y$ type properties,  H$_6$ has $\sigma_x$ like character, while L has $\sigma_z^*$ like attributes. MOs with contribution of 50\% or more in the electronic excitation are given in Figure~\ref{MOs_4}. 

\begin{figure}[hbt!]
\includegraphics[width=0.47\textwidth]{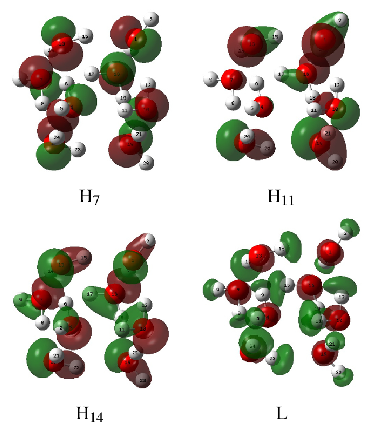}
\caption{MOs involved in first three excitations for $n=8$, with percentage contribution $\ge50\%$}
\label{MOs_8}
\end{figure}

Going on to our next most stable cluster- octamer, owing to its cubical shape as well as the degenerate energy levels, excitation of electrons from bonding H$_{14}$ ($\sigma_x$ type) to anti-bonding L (with $\sigma_z^*$ like qualitities, refer to Figure~\ref{MOs_8}) gives a notable large value of oscillator strength $=$ 0.18, which turns out to be the second highest value among all the twenty clusters. Similarly the other major peaks are from the other degenerate orbitals, namely bonding ($\sigma_x$ type) H$_{11}$ to L and non-bonding ($\pi_y$ type) H$_7$ to L with $\mathrm{f_{osc}}=0.10$ and 0.09 respectively.

For $n=12$, the highest oscillator strength value among all the twenty clusters is obtained here, with a value of 0.26 due to the bonding H$_{19}$ with $\sigma_x$ like character, which belongs to the set of degenerate orbitals and which corresponds to the excitation of  H$_{19}$~$\rightarrow$~L. The other major peaks are again obtained by the orbitals belonging to degenerate energy levels, namely H$_{1}$ and H$_{8}$ (both $\pi_y$ type) showing excitation to the LUMO state (having $\sigma_z^*$ like attributes).

When we see the optical bandgap diagram (Figure-S3 in the supplementary material), as the number of clusters goes on increasing, except for $n=3,4,8,12,16,18$ where there are blue-shifts observed due to the reasons explained above, there is a declining trend of the bandgap. It slowly decreases as the cluster size increases and reaches 7.4~eV for (H$_2$O)$_{20}$. This decrease in the energies can be explained by the number of eigenstates that get added as $n$ increases, forming a band-like structure as it approaches the bulk system, which leads to HOMO-LUMO coming closer to each other, and thus a red-shift in the eigenvalues observed. This is also the reason that in the optical spectra, in Figure~\ref{fig:optics}, for $n>10$, the excitations with higher energies show red-shift. This is comparable from the theoretical bandgap of 7.3 eV for liquid water, and with the experimentally obtained value of 6.9 eV~\cite{fang2015accurate}. A similar declining trend in the HOMO-LUMO gap has been observed in a theoretical work by Cabral do Couto \textit{et al.}~\cite{cabral2005kohn} where they have extrapolated it for $n=30$.

\section{Summary and Conclusion}
\label{summary}	
We generated several thousand structures for each of the water clusters within a size range of N~$=2-20$, using the artificial bee colony algorithm with the classical TIP4P force field. The first 10 lowest-energy structures were further optimized using an {\it ab initio} method implemented in the Gaussian code with the HCTH exchange-correlation functional and the 6-311++G(d,p) basis set to obtain the lowest equilibrium geometry. Our results show that the equilibrium geometries obtained during the structural search  are lowest in energy than the previously reported results. We reported the structural, electronic, vibrational and optical properties of these lowest equilibrium geometries. The structural stability of cluster was studied using various descriptors such as binding energy, ionization potentials, fragmentation energy, first and second energy difference, vibrational and optical spectra. We also calculated the shape deformation parameter to find the shape of these clusters. 

The binding energy per molecule shows the sharp rise up to N=10. For N$>$10, there is a slight increase in it. The energetic analysis shows that clusters with N~$=4,8,12,14,16$ and 19 are more stable. The analysis of fragmentation energies of the clusters also support these findings. The detailed analysis of the non covalent interactions~(hydrogen bonds) finds that it plays a very crucial role in stabilizing the water clusters. We also confirmed the stability of these clusters using frequency analysis from vibrational spectroscopy studies. Infrared spectroscopy reveals three distinct bands in water clusters: intermolecular O...H vibrations (23–1191 cm$^{-1}$), intramolecular H-O-H bending (1600–1741 cm$^{-1}$) and O-H stretching (3229–3877 cm$^{-1}$). The spectra show the strongest intensity for the lowest frequency symmetric stretching modes, along with a red-shift in the stretching vibrations.
        
We calculated the absorption spectra of these clusters using the B3LYP exchange-correlation functional. The optical bandgap ranges from 7.14 eV to 8.17 eV, belongs to the ultraviolet region. As expected, line broadening occurs for clusters with \( n \geq 10 \), showing an increase in spectral lines. Notably, only the stable clusters exhibit maximum oscillator strength. In all cases, the first excitation corresponds to a \( \pi \rightarrow \sigma^* \) transition.

\begin{acknowledgments}
One of the authors VKB acknowledges University of Mumbai for providing UGC fellowship and DST for Junior Research Fellowship (JRF) under DST-PURSE research program for partial funding of this work.
We are pleased to acknowledge the National Param Supercomputing Facility (NPSF), CDAC, Pune, India, for providing computing facility. Some parts of the present work have been carried out using NPSF, CDAC, Pune.
\end{acknowledgments}

\bibliographystyle{unsrtnat}
\bibliography{bibliography}

\end{document}

%% file: table1.tex
\begin{table}[hbt!]
\caption{Comparison of energies (in hartrees) of lowest energy structures of water clusters, $n=1-20$ obtained in this work, by Rakshit \textit{et al.}~\cite{rakshit2019atlas} and by others~\cite{tsai1993theoretical,sadlej1999theoretical,bulusu2006lowest,xantheas2002development}. (Rounded up to two decimals)}
\small\begin{tabular}{p{1cm}p{1.5cm}p{2.4cm}p{2.4cm}}\\\hline
$n$ & This work & Rakshit \textit{et al.}~\cite{rakshit2019atlas}  & Others~\cite{tsai1993theoretical,sadlej1999theoretical,bulusu2006lowest,xantheas2002development}  \\
\hline
1 & -76.45 & -76.33 & -   \\
2 & -152.91 & - & -152.73~\cite{xantheas2002development}    \\
3 & -229.37 & -229.00 & -229.11~\cite{xantheas2002development}   \\
4 & -305.84  & -305.36 & -305.49~\cite{xantheas2002development}  \\
5 & -382.30  & -381.70 & -381.86~\cite{xantheas2002development}  \\
6 & -458.76 & -458.05 & -458.24~\cite{xantheas2002development}   \\
7 & -535.22 & -534.40 & -533.68~\cite{sadlej1999theoretical}   \\
8 & -611.70 & -610.75 & -609.93~\cite{sadlej1999theoretical}   \\
& & & -609.70~\cite{tsai1993theoretical} \\
9 & -688.16 & -687.10 & -686.18~\cite{sadlej1999theoretical}   \\
10 & -764.63 & -763.45 & -762.42~\cite{sadlej1999theoretical}  \\
11 & -841.09  & -839.79 & -840.13~\cite{bulusu2006lowest} \\
12 & -917.56  & -916.14 & -914.57~\cite{tsai1993theoretical}   \\
13 & -994.02 & -992.49 & -992.12~\cite{bulusu2006lowest}  \\
14 & -1070.49 & -1068.84 & -  \\
15 & -1146.95 & - & -   \\
16 & -1223.42 & -1221.54 & -   \\
17 & -1299.88 & -1296.73 & -   \\
18 & -1376.35 & -1374.23 & -   \\
19 & -1452.82 & -1450.58 & -   \\
20 & -1529.28 & -1526.93 & -   \\
\hline
\end{tabular}
\label{tab:comparison}
\end{table}